\documentclass[12pt]{article}

\begin{document}

\title{\bf On the Physical Properties of Spherically Symmetric
Self-Similar Solutions}

\author{M. Sharif \thanks{e-mail: hasharif@yahoo.com} and Sehar Aziz \\
Department of Mathematics, University of the Punjab,\\
Quaid-e-Azam Campus, Lahore-54590, Pakistan.}

\date{}

\maketitle

\begin{abstract}
In this paper, we are exploring some of the properties of the
self-similar solutions of the first kind. In particular, we shall
discuss the kinematic properties and also check the singularities
of these solutions. We discuss these properties both in co-moving
and also in non co-moving (only in the radial direction)
coordinates. Some interesting features of these solutions turn up.
\end{abstract}

\newpage

\section{Introduction}

A set of field equations remains invariant under a scale
transformation if we assume appropriate matter field. This implies
the existence of scale invariant solutions to the field equations.
Such solutions are called self-similar solutions. The special
feature of self-similar solutions is that, by a suitable
coordinate transformations, the number of independent variables
can be reduced by one and hence reduces the field equations. In
other words, self-similarity refers to an invariance which simply
allows the reduction of a system of partial differential equations
to ordinary differential equations.

Similarity solutions were first studied in General Relativity (GR)
by Cahill and Taub [1]. They studied these solutions in the
cosmological context and under the assumption of spherically
symmetric distribution of a self-gravitating perfect fluid. They
assumed that the solution was such that the dependent variables
are essentially functions of a single independent variable
constructed as a dimensionless combination of the independent
variables and that the model contains no other dimensional
constants. They showed that the existence of a similarity of the
first kind in this situation could be invariantly formulated in
terms of the existence of a Homothetic vector (HV).

In GR, self-similarity is defined by the existence of a HV field.
Such similarity is called the first kind (or homothety). There
exists a natural generalization of homothety called Kinematic
self-similarity, which is defined by the existence of a kinematic
self-similar (KSS) vector field. The basic condition
characterizing a manifold vector field $\xi$ as a self-similar
generator [2] is given by
\begin{equation}
\pounds_{\xi}A=\lambda A,
\end{equation}
where $\lambda$ is constant and $A$ is independent physical field.
This field can be scalar (e.g. $\mu$), vector (e.g. $u_{a}$), or
tensor (e.g. $g_{ab}$). In GR, the gravitational field is
represented by the metric tensor $g_{ab}$, and an appropriate
definition of geometrical self-similarity is necessary.

A kinematic self-similarity satisfies the condition
\begin{equation}
\pounds_{\xi} u_{a} = \alpha u_{a},
\end{equation}
with
\begin{equation}
\pounds_{\xi} h_{ab} =2\delta h_{ab},
\end{equation}
where $\alpha$ and $\delta$ are constants and
$h_{ab}=g_{ab}-u_au_b$ is the projection tensor.

KSS perfect fluid solutions have been explored by several authors.
Carter and Henriksen [3] defined the other kinds of
self-similarity namely second, zeroth and infinite kind. In the
context of kinematic self-similarity, homothety is considered as
the first kind.

The only barotropic equation of state which is compatible with
self-similarity of first kind is $p=k\mu$. Carr [4] has classified
the self-similar perfect fluid solutions of first kind with this
equation of state for the dust case ($k=0$) and the case $0<k<1$
has been studied by Carr and Coley [5]. Coley [6] has shown that
the Friedmann Robertson Walker solution is the only spherically
symmetric homothetic perfect fluid solution in the parallel case.
McIntosh [7] has discussed that a stiff fluid ($k=1$) is the only
compatible perfect fluid with the homothety in the orthogonal
case.

Benoit and Coley [8] have studied spherically symmetric spacetimes
which admit a KSS vector of the second and zeroth kind. Sintes et
al. [9] have considered spacetimes which admit a KSS vector of
infinite kind. In all these papers the equation of state has not
been specified.

In recent papers, Maeda et al. [10,11] investigated the KSS vector
of the second kind in the tilted case. They assumed the perfect
fluid spacetime obeying a relativistic polytropic equation of
state. Further, they assumed two kinds of polytropic equation of
state in GR and showed that such spacetimes must be vacuum in both
cases. They studied the case in which a KSS vector is not only
tilted to the fluid flow but also parallel or orthogonal.

Daud and Ziad [12] have found the homotheties of spherically
symetric spacetimes admitting maximal isometry groups larger than
SO(3) along with their metrics. They have used the homothety
equations without imposing any restriction on the stress-energy
tensor. It would be worth interesting to explore the physical
properties of these solutions. In this paper, we would evaluate
the kinematical properties of these self-similar solutions of the
first kind. We shall discuss the physical properties, such as
acceleration, rotation, expansion, shear, shear invariant and
expansion rate both in co-moving and non co-moving (only in radial
direction) coordinates. Further, we would look for the
singularities of these solutions.

The paper has been organised as follows. In the next section, we
shall write all such solutions. In section 3, we shall discuss the
physical properties of these solutions of the first kind both in
co-moving and non co-moving coordinates. In section 4,
singularities of these solutions will be explored. Finally, in the
last section, we shall conclude the results.

\section{Self-Similar Spherically Symmetric Solutions}

The line element of general spherically symmetric spacetime is
given by [13]
\begin{equation}
ds^2=e^{\nu(t,r)}dt^2- e^{\lambda(t,r)}dr^2- e^{x(t,r)}d\Omega^2,
\end{equation}
where $d\Omega^2=d\theta^2+\sin^2\theta d\phi^2$.

Daud and Ziad [12] have solved homothetic equations and found the
different self-similar solutions of the first kind. There are two
classes of such solutions, one of which admit 5 HVs and the second
class admit 7 HVs. The first class of self-similar solutions which
admit 5 HVs is given by the following three different metrics
\begin{equation}
ds^2=r^{2(1-\alpha)} dt^2- dr^2- r^2 d\Omega^2,
\end{equation}
where $\alpha$ is an arbitrary constant but cannot be $1$.
\begin{equation}
ds^2= dt^2- t^{2(1-\alpha)}dr^2- t^2 d\Omega^2,
\end{equation}
\begin{equation}
ds^2= dt^2- dr^2- \beta^2(t+\alpha r)^2 d\Omega^2,
\end{equation}
where $\beta$ is an arbitrary constant. The second class of
self-similar solutions which admit 7 HVs is given by the line
element
\begin{equation}
ds^2= dt^2-  e^{x(t)}[dr^2+ \Sigma^2(m,r)d\Omega^2],
\end{equation}
where $\Sigma=a \sinh(r/a)$, $r$ and $\sin(r/a)$ for $m =1/a^2$,
$0$, $-1/a^2$ ($a$ being an arbitrary constant) respectively
subject to the following constraint
$$2me^{-x}+\ddot{x}\neq0,$$
(dot denotes differentiation with respect to $t$) and
\begin{eqnarray}
e^{x}&=&2(t\alpha+\gamma)^2,\quad\quad
\alpha^2\neq\frac{1}{a^2},\nonumber\\
&=&{\delta}(t-\gamma)^{\alpha},\quad\quad\alpha\neq 0,\nonumber\\
&=&(t\alpha +\gamma)^2,\quad\quad\alpha^2\neq -\frac{1}{a^2},
\end{eqnarray}
where $\gamma$ is an arbitrary constant. Another self-similar
solution which also admits 7 HVs is given by the following metric
\begin{equation}
ds^2=e^{y(r)}[dt^2- a^2\cosh(\frac{t}{a})d\Omega^2]-dr^2
\end{equation}
subject to the constraint
$$2e^{-y}+a^2y''\neq0,$$
(prime indicates derivative with respect to $r$) and
\begin{equation}
e^{y}=( r\alpha+\beta)^2,\quad\quad\alpha^2\neq\frac{1}{a^2}.
\end{equation}

\section{Kinematics of the Velocity Field}

In this section, we shall discuss some of the kinematical
properties of the self-similar solutions given by Eqs.(5)-(8) and
(10) both in co-moving and non co-moving coordinates. The
kinematical properties [13] can be listed as follows. The
acceleration is defined by
\begin{equation}
\dot{u}_a  = u_{a;b}u^b.
\end{equation}
The rotation is given by
\begin{equation}
\omega_{ab}=u_{[a;b]}+\dot{u}_{[a}u_{b]}.
\end{equation}
The expansion scalar, which determines the volume behaviour of the
fluid, is defined by
\begin{equation}
\Theta=u^{a}_{;a}.
\end{equation}
The shear tensor, which provides the distortion arising in fluid
flow leaving the volume invariant, can be found by
\begin{equation}
\sigma_{ab}=u_{(a;b)}+\dot{u}_{(a}u_{b)}-\frac{1}{3}\Theta h_{ab},
\end{equation}
The shear invariant is given by
\begin{equation}
\sigma=\sigma_{ab}\sigma^{ab}.
\end{equation}
The rate of change of expansion with respect to proper time is
given by Raychaudhuri's equation [14]
\begin{equation}
\frac{d\Theta}{d\tau}=-\frac{1}{3}\Theta^{2}-\sigma_{ab}\sigma^{ab}+
\omega_{ab}u^{a}u^{b}-R_{ab}u^{a}u^{b}.
\end{equation}
Now we discuss the properties of these solutions both in co-moving
and also in non co-moving coordinates.

\subsection{Kinematic Properties in Co-Moving Coordinates}

First we evaluate the kinematical properties of the self-similar
solutions in the co-moving coordinates.

For the first solution given by Eq.(5), the acceleration and
expansion are zero but one of the rotation components exists given
by
\begin{equation}
\omega_{01}=\frac{1-\alpha}{r^\alpha}.
\end{equation}
The only non-zero component of the shear is
\begin{equation}
\sigma_{01}=-\omega_{01}
\end{equation}
and consequently the shear invariant becomes
\begin{equation}
\sigma=-\frac{(1-\alpha)^2}{r^2}.
\end{equation}
By using Raychaudhuri equation, the rate of change of expansion
turns out to be
\begin{equation}
\frac{d\Theta}{d\tau}=\frac{\alpha-1}{r^2}.
\end{equation}
For the second self-similar solution given by Eq.(6), the
acceleration, rotation and expansion are zero. The non-zero shear
components are
\begin{equation}
\sigma_{11}=2(\alpha-1)t^{(1-2\alpha)},\quad\sigma_{22}=-2t,\quad
\sigma_{33}=\sigma_{22}\sin^2\theta.
\end{equation}
The shear invariant becomes
\begin{equation}
\sigma=\frac{4}{t^2}[(\alpha-1)^2+2].
\end{equation}
The expansion rate is given by Raychaudhuri equation as follows
\begin{equation}
\frac{d\Theta}{d\tau}=-\frac{1}{t^2}(5\alpha^2-9\alpha+12).
\end{equation}
Now we evaluate the above quantities for the self-similar solution
given by Eq.(7). It is easy to see that the acceleration, rotation
and expansion turn out to be zero. The non-zero components of the
shear are
\begin{equation}
\sigma_{22}=-2\beta^2(t+\alpha r),\quad
\sigma_{33}=\sigma_{22}\sin^2\theta.
\end{equation}
The shear invariant takes the following form
\begin{equation}
\sigma=\frac{8}{(t+\alpha r)^2}.
\end{equation}
The rate of change of expansion is
\begin{equation}
\frac{d\Theta}{d\tau}=-\sigma.
\end{equation}
Now we calculate the above kinematical quantities for another
self-similar solutions given by Eq.(8). For this class, the
acceleration, rotation and expansion are zero. The non-zero
components of the shear take the form
\begin{eqnarray}
\sigma_{11}&=&-x_te^{x},\nonumber\\
\sigma_{22}&=&-x_te^{x}\Sigma^2(m,r),\nonumber\\
\sigma_{33}&=&\sigma_{22}\sin^2\theta.
\end{eqnarray}
The shear invariant becomes
\begin{equation}
\sigma=3{x_{t}}^2.
\end{equation}
The expansion rate turns out to be
\begin{equation}
\frac{d\Theta}{d\tau}=\frac{3}{4}(2x_{tt}-3{x_{t}}^2).
\end{equation}
Finally, we explore the above quantities for the self-similar
solution given by Eq.(10). For this solution, the expansion is
zero while the acceleration and rotation are given by respectively
\begin{equation}
\dot{u}_1=-\frac{y_{r}}{2},
\end{equation}
\begin{equation}
\omega_{01}=y_{r}e^{\frac{y}{2}}.
\end{equation}
The non-zero components of the shear are
\begin{eqnarray}
\sigma_{01}&=&-\omega_{01},\nonumber\\
\sigma_{22}&=&-2a \cosh(\frac{t}{a})\sinh(\frac{t}{a})
e^{\frac{y}{2}},\nonumber\\
\sigma_{33}&=&\sigma_{22}\sin^2\theta.
\end{eqnarray}
The shear invariant becomes
\begin{equation}
\sigma=-{y_{r}}^2+8\frac{\tanh(\frac{t}{a})}{a^2e^{y}}.
\end{equation}
The rate of change of expansion is given by
\begin{equation}
\frac{d\Theta}{d\tau}=-\sigma+\frac{2a^2y_{rr}e^{y}
+3a^2{y_{r}}^2e^{y}-8}{4a^2e^{y}}.
\end{equation}

\subsection{Kinematic Properties in Non Co-Moving Coordinates}

Here we discuss the kinematical properties of the self-similar
solutions in the non co-moving coordinates only in radial
direction.

For the first solution, the acceleration components turn out to be
\begin{equation}
\dot{u}_0=\frac{1-\alpha}{r^\alpha},\quad
\dot{u}_1=\frac{\alpha-1}{r}.
\end{equation}
The rotation component is given by
\begin{equation}
\omega_{01}=\dot{u}_0
\end{equation}
and the expansion becomes
\begin{equation}
\Theta=\frac{\alpha-3}{r}.
\end{equation}
The non-zero components of the shear are
\begin{eqnarray}
\sigma_{00}=4(1-\alpha)r^{(1-2\alpha)},\quad
\sigma_{11}=-\frac{4\alpha}{3r},\nonumber\\
\sigma_{22}=\frac{(\alpha-9)r}{3},\quad
\sigma_{33}=\sigma_{22}\sin^2\theta,\quad
\sigma_{01}=\frac{2(4\alpha-3)}{3r^\alpha}
\end{eqnarray}
and the shear invariant is
\begin{equation}
\sigma=\frac{2(45+13\alpha^2-38\alpha)}{3r^2}.
\end{equation}
The expansion rate turns out to be
\begin{equation}
\frac{d\Theta}{d\tau}=-\frac{1}{3r^2}(27\alpha^2
-76\alpha+93)+\frac{\alpha-1}{r}.
\end{equation}
For the second solution, the acceleration components become
\begin{equation}
\dot{u}_0=\frac{\alpha-1}{t},\quad
\dot{u}_1=\frac{(1-\alpha)}{t^\alpha}.
\end{equation}
The rotation component is given by
\begin{equation}
\omega_{01}=-\dot{u}_1
\end{equation}
and the expansion becomes
\begin{equation}
\Theta=\frac{(3-\alpha)}{t}.
\end{equation}
The non-zero components of the shear are
\begin{eqnarray}
\sigma_{00}=2\dot{u}_0,\quad
\sigma_{11}=\frac{2}{3}(5\alpha-3)t^{(1-2\alpha)},\nonumber\\
\sigma_{22}=-2t,\quad \sigma_{33}=\sigma_{22}\sin^2\theta,\quad
\sigma_{01}=-\frac{2\alpha}{t^\alpha}.
\end{eqnarray}
The shear invariant becomes
\begin{equation}
\sigma=\frac{4}{t^2}(\frac{25}{9}\alpha^2-\frac{16}{3}\alpha+4).
\end{equation}
The rate of change of expansion will become
\begin{eqnarray}
\frac{d\Theta}{d\tau}&=&-\frac{1}{9t^2}(112\alpha^2-219\alpha+171)
+\frac{(1-\alpha)}{t}-\frac{(1-\alpha)(\alpha-2)}{t^\alpha}.
\end{eqnarray}
For the self-similar solution given by Eq.(7), the acceleration,
rotation and expansion are zero. The non-zero components of the
shear are
\begin{equation}
\sigma_{22}=2\beta^2(t+\alpha r)(\alpha-1),\quad
\sigma_{33}=\sigma_{22}\sin^2\theta.
\end{equation}
The shear invariant becomes
\begin{equation}
\sigma=\frac{8(1-\alpha)^2}{(t+\alpha r)^2}.
\end{equation}
The expansion rate can be found by using Raychaudhuri equation as
given below
\begin{equation}
\frac{d\Theta}{d\tau}=-\sigma.
\end{equation}
Now we consider the metric given by Eq.(8). In this case, the
acceleration components turn out to be
\begin{equation}
\dot{u}_0=-\frac{x_t}{2},\quad
\dot{u}_1=\frac{x_t}{2}e^\frac{x}{2}.
\end{equation}
The only non-zero rotation component is given by
\begin{equation}
\omega_{01}=-\dot{u}_1
\end{equation}
and the expansion becomes
\begin{equation}
\Theta=\frac{3x_t}{2}-2e^{-\frac{x}{2}}\frac{\Sigma_r}{\Sigma}.
\end{equation}
The non-zero components of the shear are
\begin{eqnarray}
\sigma_{00}=-x_t,\quad
\sigma_{11}=-2(e^{x}+\frac{2\Sigma_r}{3\Sigma}e^{\frac{x}{2}}),\quad
\sigma_{22}=-\frac{5}{3}e^{\frac{x}{2}}\Sigma
\Sigma_r,\nonumber\\
\sigma_{33}=\sigma_{22}\sin^2\theta,\quad
\sigma_{01}=x_te^{\frac{x}{2}}+\frac{2\Sigma_r}{3\Sigma}
\end{eqnarray}
and consequently the shear invariant becomes
\begin{equation}
\sigma={x_{t}}^2+\frac{4}{e^{2x}}(e^x+\frac{2\Sigma_r}
{3\Sigma}e^\frac{x}{2})^2-\frac{1}{e^x}(x_te^\frac{x}{2}
+\frac{2\Sigma_r}{3\Sigma})^2+\frac{50{\Sigma_r}^2}{9\Sigma^2e^x}.
\end{equation}
The rate of change of expansion will be
\begin{eqnarray}
\frac{d\Theta}{d\tau}=-\frac{3}{2}{x_t}^2(1+\frac{e^\frac{x}{2}}
{2})-\frac{x_{tt}}{2}(3+e^\frac{x}{2})+\frac{2\Sigma_r}{3\Sigma}
(5e^{-\frac{x}{2}}x_t-8e^{-x})\nonumber\\
-\frac{74{\Sigma_r}^2}{9e^x\Sigma^2}
+2e^{-\frac{x}{2}}\frac{\Sigma_{rr}}{\Sigma}+\frac{x_t}{2}-4.
\end{eqnarray}
Finally, we evaluate the kinematical properties of the
self-similar solution given Eq.(10). For this solution, the
acceleration components become
\begin{equation}
\dot{u}_0=\frac{y_r}{2}e^\frac{y}{2},\quad
\dot{u}_1=-\frac{y_r}{2}.
\end{equation}
The non-zero rotation component will be
\begin{equation}
\omega_{01}=\dot{u}_0
\end{equation}
and the expansion is given by
\begin{equation}
\Theta=\frac{2}{ae^\frac{y}{2}}\tanh(\frac{t}{a})-\frac{3}{2}y_r.
\end{equation}
The non-zero shear components are
\begin{eqnarray}
\sigma_{00}&=&2y_re^{y},\quad
\sigma_{11}=\frac{4\tanh(\frac{t}{a})}{3ae^\frac{y}{2}},\nonumber\\
\sigma_{22}&=&-a\cosh(\frac{t}{a})[e^\frac{y}{2}(2\sinh(\frac{t}{a})
+y_re^\frac{y}{2}a\cosh(\frac{t}{a}))\nonumber\\
&+&\frac{a}{3}\cosh(\frac{t}{a})(\frac{2}{ae^\frac{y}{2}}
\tanh(\frac{t}{a})-\frac{3}{2}y_r)],\nonumber\\
\sigma_{33}&=&\sigma_{22}\sin^2\theta,\nonumber\\
\sigma_{01}&=&-y_re^\frac{y}{2}-\frac{2}{3a}\tanh(\frac{t}{a}).
\end{eqnarray}
The shear invariant is
\begin{eqnarray}
\sigma&=&3{y_r}^2-\frac{4}{3a}y_re^{-\frac{y}{2}}\tanh(\frac{t}{a})
-\frac{4}{3a^2e^y}\tanh^2(\frac{t}{a})\nonumber\\
&+&\frac{2e^{-y}} {a^2\cosh^2(\frac{t}{a})}(2\sinh(\frac{t}{a})
+y_re^\frac{y}{2}a\cosh(\frac{t}{a}))^2\nonumber\\
&+&\frac{2e^{-2y}}{9}(\frac{2}{ae^{\frac{y}{2}}}\tanh(\frac{t}{a})
-\frac{3}{2}y_r)^2+\frac{4e^{-\frac{3y}{2}}}{3a\cosh(\frac{t}{a})}
(\frac{2}{ae^{\frac{y}{2}}}\tanh(\frac{t}{a})\nonumber\\
&-&\frac{3}{2}y_r)(2\sinh(\frac{t}{a})
+y_re^\frac{y}{2}a\cosh(\frac{t}{a})).
\end{eqnarray}
The expansion rate is given by
\begin{eqnarray}
\frac{d\Theta}{d\tau}&=&-\frac{1}{3}(\frac{2}{ae^\frac{y}{2}}
\tanh(\frac{t}{a})-\frac{3y_r}{2})^2+\frac{3}{2}y_{rr}
+\frac{3}{4}{y_r}^2\nonumber\\
&-&\frac{1}{4a^2e^y}(2y_{rr}a^2e^y
+3{y_r}^2e^ya^2-8)-\frac{y_r}{2}-\sigma.
\end{eqnarray}

\section{Singularities}

In this section, we shall discuss the singularities of the
self-similar solutions. The Kretschmann scalar is defined by
\begin{equation}
K=R_{abcd}R^{abcd},
\end{equation}
where $R_{abcd}$ is the Riemann tensor. For the first solution
given by Eq.(5), it reduces to
\begin{equation}
K= \frac{2}{r^4}(1-\alpha)[(1-\alpha)(\alpha^2+1)+1].
\end{equation}
It is obvious that K diverges at the point $r=0$. Thus the
solution is singular at $r=0$.

For the second solution given by Eq.(6), the Kretschmann scalar
turns out to be
\begin{equation}
K= \frac{1}{t^4}[(\alpha^2+2)(1-\alpha)^2+4].
\end{equation}
and hence the metric is scalar polynomial singular along $t=0.$

The Kretschmann scalar for the third solution takes the form
\begin{eqnarray}
K&=&\frac{1}{\beta^4(t+r\alpha)^8}[2t^2+2\alpha tr+\alpha^2r^2
+2\alpha\beta^2 tr+\alpha^2\beta^2r^2\nonumber\\
&-&\alpha^2\beta^2t^2-2\alpha^3\beta^2tr-\alpha^4\beta^2r^2].
\end{eqnarray}
We see that $K$ diverges only when $t=0,~r=0$ and consequently
this solution is singular at these points simultaneously.

For the class of solutions given by Eq.(9), $K$ becomes
\begin{eqnarray}
K&=&2[\frac{3}{4}(x_{tt}+\frac{1}{2}{x_t}^2)^2+\frac{2}{e^{2x}
\Sigma^2}(\Sigma_{rr}-\frac{1}{4}{x_t}^2e^{x}\Sigma)^2\nonumber\\
&+&\frac{1}{e^{2x}\Sigma^4}(1+\frac{1}{4}{x_t}^2e^{x}\Sigma^2
-{\Sigma_{r}}^2)^2].
\end{eqnarray}
This is singular only when $\Sigma=0$, i.e., when $r=0.$ Finally,
for the last solution, the Kretschmann scalar reduces to
\begin{eqnarray}
K&=&2[\frac{3}{4}(y_{rr}+\frac{1}{2}{y_r}^2)^2
+\frac{2}{a^4e^{2y}}(1-\frac{1}{4}{y_r}^2a^2e^{y})^2\nonumber\\
&+&\frac{1}{a^4e^{2y}\cosh^4(t/a)}(1+\sinh^2(t/a)
-\frac{1}{4}{y_r}^2a^2e^{y}\cosh^2(t/a))^2].
\end{eqnarray}
This does not provide any singularity and hence the last solution
is singularity free.

\section{Conclusion}

In GR, the term self-similarity can be used in two different ways.
One is for the properties of spacetimes and the other is for the
properties of matter fields. In general, these are not equivalent.
Although homothetic solutions can contain many interesting matter
fields, but those compatible with homothety are not so common. In
this paper, we have considered some self-similar solutions of the
first kind and have evaluated some physical properties of these
solutions both in co-moving and also in non co-moving coordinates.
We also have checked singularities of each solution (if any).

First we discuss the physical properties in co-moving coordinates.
For the first self-similar solution, the acceleration and the
expansion turn out to be zero while the remaining physical
quantities exist and are finite. It can be seen from Eqs.(18)-(20)
that these quantities cannot be made zero as $\alpha\neq1$.
Further all these quantities become infinite at $r=0$ and also the
expansion rate can be positive or negative according to the choice
of $\alpha$.

For the second solution, the acceleration, rotation and expansion
are zero while the rest of the quantities do not vanish as
$\alpha\neq1$. We see from Eq.(24) that the expansion rate does
provide positive/negative value according to the choice of
$\alpha$. Also, we see that shear invariant and expansion rate
turn out to be infinite for $t=0$. The third solution also gives
acceleration, rotation and expansion to be zero. In this case the
expansion rate is always negative and become infinite only at
$t=0,~r=0$. For the fourth solution again acceleration, rotation
and expansion turn out to be zero. Further, the expansion rate is
zero only if $x=constant$ otherwise it is finite. For the last
self-similar solution, only expansion is zero while the remaining
quantities are non-zero and finite.

In non co-moving coordinates, the kinematical quantities turn out
to be non-zero for all the solutions except for the solution given
by Eq.(7). The reason is that we are also considering velocity in
radial direction in addition to the temporal component. Also, we
see that the results are more complicated in the non co-moving
coordinates as compared to the co-moving coordinates.

Finally, we discuss the singularities of these solutions. We see
that the first solution becomes singular at $r=0$ while the second
solution is singular at $t=0$. The third solution turns out to be
singular only if $t=0,~r=0$ whereas the fourth solution is
singular at $r=0$. The last solution turns out to be singularity
free solution.

\newpage

\begin{description}
\item  {\bf Acknowledgment}
\end{description}

The authors acknowledge the enabling role of the Higher Education
Commission Islamabad, Pakistan and appreciate its financial
support through {\it Merit Scholarship Scheme for Ph.D. Studies in
Science and Technology (200 Scholarships)}.

\vspace{2cm}

{\bf \large References}

\begin{description}

\item{[1]} Cahill, M. E. and Taub, A.H.: Commun. Math. Phys.\textbf{21}(1971)1.

\item{[2]} Carter, B. and Henriksen, R.N.: Ann. de Phys. {\bf 14}(1989)47.

\item{[3]} Carter, B. and Henriksen, R. N.: J. Math. Phys. \textbf{32}(1991)2580.

\item{[4]} Carr, B.J.: Phys. Rev. {\bf D62}(2000)044022.

\item{[5]} Carr, B.J. and Coley, A.A.: Phys. Rev. {\bf D62}(2000)044023.

\item{[6]} Coley, A.A.: Class. Quantum. Grav. {\bf 14}(1997)87.

\item{[7]} McIntosh, C.B.G.: Gen. Relat. Gravit. {\bf 7}(1975)199.

\item{[8]} Benoit, P.M. and Coley, A.A.: Class. Quantum. Grav. {\bf 15}(1998)2397.

\item{[9]} Sintes, A.M., Benoit, P.M. and Coley, A.A.: Gen. Rel. Gravi.
{\bf 33}(2001)1863.

\item{[10]} Maeda, H., Harada, T., Iguchi, H. and Okuyama, N.: Phys. Rev.
{\bf D66}(2002)027501.

\item{[11]} Maeda, H., Harada, T., Iguchi, H. and Okuyama, N.: Prog. Theor. Phys.
{\bf 108}(2002)819; Prog. Theor. Phys. {\bf 110}(2003)25.

\item{[12]} Ahmad, Daud and Ziad, M.: J. Math. Phys. \textbf{38}(1997)2547.

\item{[13]} Stephani, H., Kramer, D., MacCallum, M.A.H.,
Hoenselaers, C. and Hearlt, E.: {\it Exact Solutions of Einstein's
Field Equations} (Cambridge University Press, 2003).

\item{[14]} Wald, R. M.: \textit{General Relativity} (University of
Chichago, Chichago, 1984).

\end{description}

\end{document}